\documentclass{aastex}%
\usepackage{amssymb}
\usepackage{amsmath}
\usepackage{amsfonts}
\usepackage{graphicx}%
\begin{document}
\bigskip\ \ \ \ \ \ \ \ {\LARGE On differences between the \textbf{B- }and
\textbf{E-}approaches and the implications for the Solar atmosphere\textbf{\ }}

\ \ \ \ \ \ \ \ \ \ \ \ \ \ \ \ \ \ \ \ \ \ \ \ \ \ \ \ \ \ \ \ \ \ \ \ \ \ \ \ \ {\large Anatoly
K. Nekrasov}

Institute of Physics of the Earth, Russian Academy of Sciences, 123995 Moscow,
Russia; anekrasov@ifz.ru; nekrasov.anatoly@gmail.com

\bigskip

\textbf{Abstract. }A simple collisional three-component plasma model
consisting of electrons, ions, and neutrals with arbitrary collision
frequencies and dynamic time scales is considered. It is shown that the usual
MHD-approach dealing with magnetic field perturbations can give other results
than the approach in which all perturbations are expressed via the perturbed
electric field. For the partially ionized plasma with strong collisional
coupling of neutrals with ions, magnetosonic (nondamping) and Alfv\'{e}n
(weakly damping) waves modified by the presence of neutrals are obtained. It
is shown that the magnetic diffusivity for Alfv\'{e}n waves appears only due
to the longitudinal current connected with the field $E_{1z}$ at the angular
propagation of perturbations relatively to the background magnetic field.

The model can be applied to different parts of a solar atmosphere and prominences.

\bigskip


\bigskip

\section{Introduction}

It is known that analytical investigations of the fluid plasma systems are
generally done by two methods. In the first one, the electric field
perturbation is excluded from equations of motion and corresponding perturbed
velocities are expressed through the magnetic field perturbation (e.g.,
Chandrasekhar 1961). This method can be called the $\mathbf{B}$-approach. It
is mainly used in astrophysics, for some problems of solar physics and
geophysics when studying magnetohydrodynamic (MHD) waves. In the second method
or in the $\mathbf{E}$-approach, the perturbed velocities of species are
expressed through the components of the electric field perturbation (e.g.,
Mikhailovskii 1975). Such an approach is mainly applied in the theory of
plasma physics, in solar physics, and in geophysics also.

However, the $\mathbf{B}$- and $\mathbf{E}$-approaches can lead to different
results. For instance, in a resistive fully-ionized electron-ion medium,
dispersion relations for MHD waves obtained via the $\mathbf{B}$-approach are
given by (e.g., Nekrasov 2009)%

\begin{equation}
\omega^{2}+i\omega\eta k^{2}-k^{2}c_{A}^{2}=0
\end{equation}
for magnetosonic waves and%

\begin{equation}
\omega^{2}+i\omega\eta k^{2}-k_{z}^{2}c_{A}^{2}=0
\end{equation}
for Alfv\'{e}n waves. Here, $\omega$ is the frequency, $k^{2}=k_{\perp}%
^{2}+k_{z}^{2}$, $k$ is the wavenumber, $c_{A}$ is the Alfv\'{e}n velocity,
$\eta$ is the magnetic diffusivity, the background magnetic field is assumed
to be directed along the axis $\mathbf{z}$, subscripts $\perp$ and $z$ denote
directions relatively to the magnetic field. We can conclude from Equations
(1) and (2) that

1. Magnetosonic and Alfv\'{e}n waves are damped due to the resistivity.

2. The resistivity is isotropic.

At the same time, the corresponding dispersion relation derived through the
$\mathbf{E}$-approach has the form (Nekrasov 2009)
\begin{equation}
\left(  \omega^{2}-k^{2}c_{A}^{2}\right)  \left(  \omega^{2}+i\omega\eta
k_{\perp}^{2}-k_{z}^{2}c_{A}^{2}\right)  =0.
\end{equation}
We see from Equation (3) that

1. Magnetosonic waves are not damped.

2. Alfv\'{e}n waves are damped due to the resistivity. The resistivity is
anisotropic and proportional to $k_{\perp}^{2}$. When $k_{\perp}=0$,
Alfv\'{e}n waves are also not damped.

These results differ from the ones obtained from Equations (1) and (2).

The real part of the frequency $\omega$ found from Equations (1) or (2), for
example, is equal to zero at the cut-off wavenumber $k_{zc}=\pm2c_{A}/\eta$ in
the case of the longitudinal propagation ($k_{\perp}=0$) (Chandrasekhar 1961;
Zaqarashvili et al. 2012 and references therein). Zaqarashvili et al. (2012)
have shown that the appearance of the cut-off wavenumber is due to some
simplifications of the basic equations. They have considered partially ionized
plasmas of the solar atmosphere in the two-fluid description, where one
component is the charged fluid (electrons and ions) and the other component is
the neutral gas. It has been shown that for the time scales longer then the
ion-neutral collision time and neglecting the corresponding Hall term one
comes to the usual single-fluid MHD equations giving the magnetosonic and
Alfv\'{e}n waves (1) and (2), respectively. Without introducing one of these
two simplifications, the cut-off wavenumber is absent.

In their study, Zaqarashvili et al. (2012) have used the $\mathbf{B}%
$-approach. However, as we have seen above, the $\mathbf{E}$-approach can give
another results. From our point of view, Equation (3) takes into account the
physical mechanism of the collisional damping correctly (see below).
Therefore, we consider here the three-component plasma consisting of
electrons, ions, and neutrals by making of use the $\mathbf{E}$-approach in a
general form, where the frequencies of collisions between different species
are arbitrary. We derive a dispersion relation for perturbations without any
simplifications and consider it in a particular case suitable for solar prominences.

The paper is organized as follows. In Section 2, we give the main equations
and find the perturbed velocities in a general form. The components of
perturbed current are obtained in Section 3. The wave equation is considered
in Section 4. In section 5, the dispersion relations for the longitudinal and
angular propagations in the case of strong collisional coupling of neutrals
with ions are derived. An applicability of obtained results to the solar
atmosphere is discussed in Section 6. In Section 7, we present conclusive remarks.

\bigskip

\section{Basic equations and expressions for perturbed velocities}

The equations of motion for species that we consider are the following:%
\begin{equation}
\frac{\partial\mathbf{v}_{e}}{\partial t}+\mathbf{v}_{e}\cdot\mathbf{\nabla
v}_{e}=\frac{q_{e}}{m_{e}}\mathbf{E+}\frac{q_{e}}{m_{e}c}\mathbf{v}_{e}%
\times\mathbf{B-}\nu_{ei}\left(  \mathbf{v}_{e}-\mathbf{v}_{i}\right)
\mathbf{-}\nu_{en}\left(  \mathbf{v}_{e}-\mathbf{v}_{n}\right)  ,
\end{equation}%
\begin{equation}
\frac{\partial\mathbf{v}_{i}}{\partial t}+\mathbf{v}_{i}\cdot\mathbf{\nabla
v}_{i}=\frac{q_{i}}{m_{i}}\mathbf{E+}\frac{q_{i}}{m_{i}c}\mathbf{v}_{i}%
\times\mathbf{B-}\nu_{ie}\left(  \mathbf{v}_{i}-\mathbf{v}_{e}\right)
\mathbf{-}\nu_{in}\left(  \mathbf{v}_{i}-\mathbf{v}_{n}\right)  ,
\end{equation}%
\begin{equation}
\frac{\partial\mathbf{v}_{n}}{\partial t}+\mathbf{v}_{n}\cdot\mathbf{\nabla
v}_{n}=-\nu_{ne}\left(  \mathbf{v}_{n}-\mathbf{v}_{e}\right)  -\nu_{ni}\left(
\mathbf{v}_{n}-\mathbf{v}_{i}\right)  .
\end{equation}
Here, $\mathbf{v}_{j}$ is the velocity of species $j$, where $j=e,i,n$ denotes
electrons, ions, and neutrals, respectively, $q_{j}$ is the charge, $\nu_{ab}$
is the collision frequency of particle $a$ with particles $b$, $\mathbf{E}$
and $\mathbf{B}$ are the electric and magnetic fields, and $c$ is the speed of
light in vacuum. For simplicity, we use here the same momentum equations as
given in (Zaqarashvili et al. 2012) to pay attention on collisional
interactions in partially ionized plasmas in the $\mathbf{E}$-approach. We
don't take into account temperatures, viscosity, gravity etc. The plasma and
the background magnetic field are assumed to be homogeneous.

The dynamics of neutrals is only determined by collisions with charged
particles. After linearization of Equations (4)-(6) and substitution of the
perturbed velocity of neutrals $v_{n1}$ (the subject $1$ here and below
denotes the perturbed values) into the linearized equations (4) and (5), we
obtain the equation for $\mathbf{v}_{j1}$, $j=e,i$,
\begin{equation}
\beta_{j}\frac{\partial\mathbf{v}_{j1}}{\partial t}=\mathbf{F}_{j1}%
\mathbf{+}\frac{q_{j}}{m_{j}c}\mathbf{v}_{j1}\times\mathbf{B}_{0}.
\end{equation}
Here,
\begin{align}
\mathbf{F}_{e1}  &  =\frac{q_{e}}{m_{e}}\mathbf{E}_{1}-\alpha_{e}\left(
\mathbf{v}_{e1}-\mathbf{v}_{i1}\right)  ,\\
\mathbf{F}_{i1}  &  =\frac{q_{i}}{m_{i}}\mathbf{E}_{1}-\alpha_{i}\left(
\mathbf{v}_{i1}-\mathbf{v}_{e1}\right) \nonumber
\end{align}
and%
\begin{align}
\beta_{j}  &  =1+\frac{\nu_{jn0}}{\alpha_{n}},\\
\alpha_{e}  &  =\nu_{ei0}+\frac{\nu_{en0}\nu_{ni0}}{\alpha_{n}},\nonumber\\
\alpha_{i}  &  =\nu_{ie0}+\frac{\nu_{in0}\nu_{ne0}}{\alpha_{n}},\nonumber\\
\alpha_{n}  &  =\frac{\partial}{\partial t}+\left(  \nu_{ne0}+\nu
_{ni0}\right)  .\nonumber
\end{align}
The subscript $0$ denotes unperturbed collision frequencies. Solution of
Equation (7) is given by%
\begin{align}
\Omega_{j}^{2}v_{j1x}  &  =\omega_{cj}F_{j1y}+\beta_{j}\frac{\partial F_{j1x}%
}{\partial t},\\
\Omega_{j}^{2}v_{j1y}  &  =-\omega_{cj}F_{j1x}+\beta_{j}\frac{\partial
F_{j1y}}{\partial t},\nonumber\\
\beta_{j}\frac{\partial v_{j1z}}{\partial t}  &  =F_{j1z},\nonumber
\end{align}
where%
\begin{equation}
\Omega_{j}^{2}=\beta_{j}^{2}\frac{\partial^{2}}{\partial t^{2}}+\omega
_{cj}^{2}%
\end{equation}
and $\omega_{cj}=q_{j}B_{0}/m_{j}c$ is the cyclotron frequency. The background
magnetic field $\mathbf{B}_{0}$ is directed along the axis $\mathbf{z}$.

\bigskip

\section{The perturbed current}

Let us now find the perturbed current $\mathbf{j}_{1}\mathbf{=}\sum_{j}%
q_{j}n_{j0}\mathbf{v}_{j1}=q_{i}n_{0}\left(  \mathbf{v}_{i1}-\mathbf{v}%
_{e1}\right)  $, where we have assumed the condition of quasineutrality
$n_{0e}=n_{0i}=n_{0}$ ($q_{e}=-q_{i}$). Using Equation (10) for the transverse
velocity and taking into account Equation (8), we obtain two equations
\begin{align}
aj_{1x}+bj_{1y}  &  =\frac{q_{i}^{2}n_{0}}{m_{i}}\left(  dE_{1y}%
+f\frac{\partial E_{1x}}{\partial t}\right)  ,\\
aj_{1y}-bj_{1x}  &  =\frac{q_{i}^{2}n_{0}}{m_{i}}\left(  -dE_{1x}%
+f\frac{\partial E_{1y}}{\partial t}\right)  ,\nonumber
\end{align}
where notations are introduced%
\begin{align}
a  &  =1+\left(  \frac{\alpha_{i}\beta_{i}}{\Omega_{i}^{2}}+\frac{\alpha
_{e}\beta_{e}}{\Omega_{e}^{2}}\right)  \frac{\partial}{\partial t},\\
b  &  =\frac{\omega_{ci}\alpha_{i}}{\Omega_{i}^{2}}+\frac{\omega_{ce}%
\alpha_{e}}{\Omega_{e}^{2}},\nonumber\\
d  &  =\frac{\omega_{ci}}{\Omega_{i}^{2}}+\frac{\omega_{ce}}{\Omega_{e}^{2}%
}\frac{m_{i}}{m_{e}},\nonumber\\
f  &  =\frac{\beta_{i}}{\Omega_{i}^{2}}+\frac{\beta_{e}}{\Omega_{e}^{2}}%
\frac{m_{i}}{m_{e}}.\nonumber
\end{align}
It is convenient to find a solution of Equation (12) for the value
$4\pi\left(  \partial/\partial t\right)  ^{-1}j_{1x,y}$. Then we obtain%
\begin{align}
4\pi\left(  \frac{\partial}{\partial t}\right)  ^{-1}j_{1x}  &  =\varepsilon
_{xx}E_{1x}+\varepsilon_{xy}E_{1y},\\
4\pi\left(  \frac{\partial}{\partial t}\right)  ^{-1}j_{1y}  &  =-\varepsilon
_{xy}E_{1x}+\varepsilon_{xx}E_{1y},\nonumber
\end{align}
where%
\begin{align}
\varepsilon_{xx}  &  =\frac{\omega_{pi}^{2}}{\left(  a^{2}+b^{2}\right)
}\left[  bd\left(  \frac{\partial}{\partial t}\right)  ^{-1}+af\right]  ,\\
\varepsilon_{xy}  &  =\frac{\omega_{pi}^{2}}{\left(  a^{2}+b^{2}\right)
}\left[  ad\left(  \frac{\partial}{\partial t}\right)  ^{-1}-bf\right]
\nonumber
\end{align}
and $\omega_{pi}=\left(  4\pi q_{i}^{2}n_{i0}/m_{i}\right)  ^{1/2}$ is the ion
plasma frequency.

From Equations (8) and (10), we find further the perturbed longitudinal
current $j_{1z}=q_{i}n_{0}\left(  v_{i1z}-v_{e1z}\right)  $. Calculations show
that%
\begin{equation}
4\pi\left(  \frac{\partial}{\partial t}\right)  ^{-1}j_{1z}=\varepsilon
_{zz}E_{1z},
\end{equation}
where%
\begin{equation}
\varepsilon_{zz}=\omega_{pi}^{2}\left(  \frac{\partial}{\partial t}%
+\frac{\alpha_{i}}{\beta_{i}}+\frac{\alpha_{e}}{\beta_{e}}\right)
^{-1}\left(  \frac{1}{\beta_{i}}+\frac{1}{\beta_{e}}\frac{m_{i}}{m_{e}%
}\right)  \left(  \frac{\partial}{\partial t}\right)  ^{-1}.
\end{equation}

\bigskip

\section{Wave equation}

Our model is azimuthally symmetrical. Therefore, we can set $\partial/\partial
x=0$. Then, from Faraday's%
\[
\mathbf{\nabla\times E=-}\frac{1}{c}\frac{\partial\mathbf{B}}{\partial t}
\]
and Ampere`s%
\[
\mathbf{\nabla\times B=}\frac{4\pi}{c}\mathbf{j}
\]
laws, we can obtain wave equations for perturbations. Using Equations (14) and
(16), we find%
\begin{align}
\left(  n^{2}-\varepsilon_{xx}\right)  E_{1x}-\varepsilon_{xy}E_{1y}  &  =0,\\
\varepsilon_{xy}E_{1x}+\left(  n_{z}^{2}-\varepsilon_{xx}\right)  E_{1y}%
-n_{y}n_{z}E_{1z}  &  =0,\nonumber\\
-n_{y}n_{z}E_{1y}+\left(  n_{y}^{2}-\varepsilon_{zz}\right)  E_{1z}  &
=0,\nonumber
\end{align}
where $n^{2}=n_{y}^{2}+n_{z}^{2}$, $n_{y,z}=c\left(  \partial/\partial
y,z\right)  \left(  \partial/\partial t\right)  ^{-1}$.

Components $\varepsilon_{xx}$, $\varepsilon_{xy}$, and $\varepsilon_{zz}$
given by Equations (15) and (17) have a general form and can be applied at
arbitrary correlations between collision frequencies of species and dynamic
time scales. Therefore, it is possible to study wave propagation in partially
ionized plasmas in different regions, for example, of the solar atmosphere.
Below, we consider one specific case.

\bigskip

\section{Dispersion relation}

For perturbations of the form $\mathbf{E}_{1}\propto\mathbf{E}_{1\mathbf{k}%
}\exp\left(  i\mathbf{k\cdot r-}i\omega t\right)  $ Equation (18) becomes
algebraic and the determinant of this system gives a dispersion relation in a
general form.

\bigskip

\subsection{Longitudinal propagation}

We first consider waves propagating almost along the background magnetic field
when $k_{y}\approx0$. The case $k_{y}=0$ was treated by Zaqarashvili et al.
(2012). Then the dispersion relation is the following:%
\begin{equation}
\left(  n_{z}^{2}-\varepsilon_{xx}\right)  =\pm i\varepsilon_{xy}.
\end{equation}
This equation describes, as it is well-known, two circularly-polarized waves:
the magnetosonic and Alfv\'{e}n waves rotating in opposite directions in the
plane perpendicular to the magnetic field $\mathbf{B}_{0}$. The main condition
for the value $k_{y}^{2}$ can be found from $n_{y}^{2}\ll\varepsilon_{xy}$. \ 

To find the values $\varepsilon_{xx}$ and $\varepsilon_{xy}$, we must specify
collision frequencies and time scales. We will consider the case in which
\begin{equation}
\omega\ll\nu_{ni0}%
\end{equation}
when there is a strong collisional coupling of neutrals with ions. Then, we
obtain (see Equation (9)) $\alpha_{n}=\nu_{ni0}$, $\beta_{e}=1+\nu_{en0}%
/\nu_{ni0}$, $\beta_{i}=1+\rho_{n0}/\rho_{i0}$, $\alpha_{e}=\nu_{ei0}%
+\nu_{en0}$, $\alpha_{i}=\left(  m_{e}/m_{i}\right)  \alpha_{e}$, where
$\rho_{a0}=m_{a}n_{a0}$. Further, we assume the condition of magnetization
(see Equation (11))%
\begin{equation}
\omega_{ci}^{2}\gg\beta_{i}^{2}\omega^{2},
\end{equation}
which is easily satisfied. We note that according to Equation (21) the
electrons are also magnetized $\left(  \omega_{ce}^{2}\gg\beta_{e}^{2}%
\omega^{2}\right)  $ because $\nu_{ni0}\gg\nu_{ne0}$. Under these conditions,
the values given by Equation (13) are the following:%
\begin{equation}
a\simeq1,b=\frac{\alpha_{i}\beta_{i}^{2}\omega^{2}}{\omega_{ci}^{3}}%
\ll1,d=\frac{\beta_{i}^{2}\omega^{2}}{\omega_{ci}^{3}},f=\frac{\beta_{i}%
}{\omega_{ci}^{2}}.
\end{equation}
When calculating the value $a$, we have assumed the additional condition
\begin{equation}
\omega_{ci}^{2}\gg\alpha_{i}\beta_{i}\omega.
\end{equation}
Substituting Equation (22) into Equation (15), we find $\varepsilon_{xx}%
=c^{2}/c_{A}^{2}$, $\varepsilon_{xy}=i\varepsilon_{xx}\beta_{i}\omega
/\omega_{ci}$, where $c_{A}=\left[  B_{0}^{2}/4\pi\left(  \rho_{i0}+\rho
_{n0}\right)  \right]  ^{1/2}$ is the Alfv\'{e}n velocity.

The dispersion relation (19) takes the form%
\begin{equation}
\omega^{2}\pm\beta_{i}\frac{\omega^{3}}{\omega_{ci}}-k_{z}^{2}c_{A}^{2}=0.
\end{equation}
The second term on the left-hand side of Equation (24) appearing due to the
Hall effect (a sum of the ion and neutral inertia) and describing the
dispersion is small (see Eq. (21)). We have obtained the usual equation for
Alfv\'{e}n (magnetosonic) waves modified by the presence of strong neutral-ion
collisions. These waves are not damped and have no the cut-off wavenumber.

\bigskip

\subsection{Angular propagation}

We now consider the case in which $k_{y}\neq0$. In the region $n_{y}^{2}%
\ll\varepsilon_{zz}$, we can neglect the contribution of $E_{1z}$ in Equation
(18). Assuming condition $n_{y}^{2}\gg\varepsilon_{xy}$, which is opposite to
that in Section 6, the terms $\varepsilon_{xy}E_{1x,y}$ can also be omitted.
Then, we obtain
\begin{align}
\left(  n^{2}-\varepsilon_{xx}\right)  E_{1x}  &  =0,\\
\left(  n_{z}^{2}-\varepsilon_{xx}\right)  E_{1y}  &  =0.\nonumber
\end{align}
Equation (25) describes the independent linearly-polarized magnetosonic,
$E_{1x}\neq0$, and Alfv\'{e}n, $E_{1y}\neq0$, waves. Taking into account the
field $E_{1z}$ for the Alfv\'{e}n wave (two last equalities in Equation (18)),
we find the dispersion relation%
\begin{equation}
n_{z}^{2}-\varepsilon_{xx}+n_{y}^{2}\frac{\varepsilon_{xx}}{\varepsilon_{zz}%
}=0,
\end{equation}
where $\varepsilon_{zz}$ given by Equation (17) has the form%
\begin{equation}
\varepsilon_{zz}=-\frac{\omega_{pe}^{2}}{\omega\left(  \omega\beta_{e}%
+i\alpha_{e}\right)  },
\end{equation}
where $\omega_{pe}$ is the electron plasma frequency. We note that the ions
don't contribute to the longitudinal current in our model without the thermal pressure.

We further consider the low-frequency case in which%
\begin{equation}
\omega\beta_{e}\ll\alpha_{e}.
\end{equation}
Substituting Equation (27) into Equation (26), we obtain%
\begin{equation}
\omega^{2}+i\eta_{m}k_{y}^{2}\omega-k_{z}^{2}c_{A}^{2}=0,
\end{equation}
where $\eta_{m}=$ $c^{2}\left(  \nu_{ei0}+\nu_{en0}\right)  /\omega_{pe}^{2}$
is the magnetic diffusivity modified by the electron-neutral collisions. This
Equation is analogous to Equation (3). Formally, we see that for given $k_{z}$
the real part of the frequency $\omega_{r}$ becomes zero at $k_{yc}^{2}%
=\pm2k_{z}c_{A}/\eta_{m}$. However, it is not the cut-off wavenumber in the
sense of paper by Zaqarashvili et al. (2012). We emphasize that the
contribution of the magnetic diffusivity to Equation (29) appears only due to
the longitudinal current connected with the field $E_{1z}$.

\bigskip

\section{Discussion}

The solar atmosphere, including prominences, is only partially ionized. We now
discuss the applicability conditions used in Section 5 to this medium. The
main condition for Equations (24) and (29) is the strong collision coupling
between neutrals and ions given by Equation (20). To find $\nu_{ni}$, we
consider parameters corresponding to solar quiescent prominences:
$n_{i}=10^{10}$ cm$^{-3}$, $n_{n}=2\times10^{10}$ cm$^{-3}$, and $T=8000$ K,
where subscripts $i=p$ and $n$ denote protons and neutrons, respectively
(e.g., Zaqarashvili et al. 2012). Then, using the collisional proton-neutron
cross section $\sigma_{in}=5\times10^{-15}$ cm$^{2}$ (D\'{\i}az et al. 2012),
we find (Braginskii 1965)%
\[
\nu_{ni}=n_{i}\frac{8}{3}\left(  \frac{1}{\pi}\frac{T}{m_{i}}\right)
^{1/2}\sigma_{in}\simeq61\text{ s}^{-1}.
\]
The observed periods for prominence oscillations are in the range between $30$
s (Balthasar et al. 1993) and $10-30$ hr (Foullon et al. 2009). Thus, we have
the strong neutron-proton collisional coupling. \ 

The second condition to obtain Equation (29) is Equation (28). Calculating
$\nu_{en0}$ and $\nu_{ei0}$ (Braginskii 1965), we obtain $\nu_{en0}%
=1.84\times10^{3}$ s$^{-1}$ and $\nu_{ei0}=5.75\times10^{5}$ s$^{-1} $, or
$\alpha_{e}\simeq\nu_{ei0}$ and $\beta_{e}\simeq31.16$. Thus, Equation (28) is
also satisfied.

For the magnetic field in solar prominences $B_{0}=10$ G, we have $\omega
_{ce}=1.76\times10^{8}$ s$^{-1}$ and $\omega_{ci}=\omega_{cp}=0.96\times
10^{5}$ s$^{-1}$. We see that the condition of magnetization given by Equation
(21) and the additional condition defined by Equation (23) are wittingly satisfied.

For parameters given above, the diffusivity $\eta_{m}$ is equal to $\eta
_{m}\ =1.63\times10^{7}$ cm$^{2}$ s$^{-1}$. Then, the dissipation term in
Equation (29) is much less than $k_{z}^{2}c_{A}^{2}$, where $c_{A}%
=1.26\times10^{7}$ cm s$^{-1}$, in the case $k_{z}\gg1.29k_{y}^{2}$ cm. If
formally set $k_{z}\sim k_{y}$, we obtain $\lambda_{y}\gg8$ cm that is, of
course, satisfied. Thus, this wave is the weakly damping Alfv\'{e}n one.

\bigskip

\section{Conclusion}

We have considered a simple collisional three-component plasma model
consisting of electrons, ions, and neutrals, in which collision frequencies
between different species are arbitrary. This model can be applied to
different parts of the solar atmosphere and prominences. One of the main
purpose of the paper was to show that the usual MHD-approach dealing with the
magnetic field perturbations can give other results than the approach in which
all the perturbations are expressed via the perturbed electric field. For the
partially ionized plasma of solar prominences with strong collisional coupling
of neutrals with ions, we have obtained magnetosonic (nondamping) and
Alfv\'{e}n (weakly damping) waves modified by the presence of neutrals. We
have shown that the magnetic diffusivity for Alfv\'{e}n waves appears only due
to the longitudinal current connected with the field $E_{1z}$ in the case of
angular propagation of perturbations relatively to the background magnetic field.

The values $\varepsilon_{xx}$, $\varepsilon_{xy}$, and $\varepsilon_{zz}$
given by Equations (15) and (17) have a general form and can be applied to
arbitrary correlations between collision frequencies of species and dynamic
time scales. Therefore, it is possible to study wave propagation in partially
ionized plasmas in different regimes.

The results obtained are useful for an investigation of the solar atmosphere
and other collisional astrophysical media.

\bigskip

\section{References}

Balthasar, H., Wiehr, E., Schleicher, H., \& Wohl, H. 1993, A\&A, 277, 635

Braginskii, S. I. 1965, Rev. Plasma Phys., 1, 205

D\'{\i}az, A. J., Soler, R., \& Ballester, J. L. 2012, ApJ, 754, 41

Foullon, C., Verwichte, E., \& Nakariakov, V. M. 2009, ApJ, 700, 1658

Mikhailovskii, A. B. 1974, Theory of plasma instabilities (Springer: Verlag)

Nekrasov, A. K. 2009, ApJ, 704, 80

Chandrasekhar, S. 1961, Hydrodynamic and Hydromagnetic Stability (London:
Oxford University Press)

Zaqarashvili, T. V., Carbonell, M., Ballester, J. l., \& Khodachenko, M. L.
2012, A\&A, 544, A143

\bigskip

\end{document}